\documentclass[12pt]{article}
\usepackage{graphicx}
\usepackage{amsfonts,amssymb}
\usepackage{amsmath}
\usepackage{citehack,cite}

\newcommand{\ve}[1]{{\bf #1}}
\newcommand{\vc}[1]{{\bf #1}}
\newcommand{\be}{\begin{equation}}
\newcommand{\ee}{\end{equation}}
\newcommand{\bea}{\begin{eqnarray}}
\newcommand{\eea}{\end{eqnarray}}
\newcommand{\sli}{\sum\limits}

\newcommand{\vl}{\vc{l}}
\newcommand{\lp}{\left (}
\newcommand{\rp}{\right )}

\newcommand{\rd}{\right .}

\begin{document}

\begin{center}
{\bf MORSE FLUIDS IN THE IMMEDIATE VICINITY OF THE CRITICAL POINT:
CALCULATION OF THERMODYNAMIC COEFFICIENTS} 
\end{center}

\begin{center}
{\sc I.V. Pylyuk$^{a,}$\footnote[1]{e-mail: piv@icmp.lviv.ua},
M.P. Kozlovskii$^a$, O.A. Dobush$^a$, M.V. Dufanets$^b$}
\end{center}

\begin{center}
{\it $^a$Institute for Condensed Matter Physics
of the National Academy of Sciences of Ukraine,
1~Svientsitskii Str., 79011 Lviv, Ukraine \\
$^b$Ivan Franko National University of Lviv,
50~Drahomanov Str., 79005 Lviv, Ukraine}
\end{center}

\vspace{0.5cm}

{\small
The previously proposed approach for the microscopic description of the critical behavior of Morse liquids based on the cell fluid model is applied to the case where the parameters of the Morse interaction potential correspond to alkali metals (sodium and potassium).
The critical temperatures, densities, and pressures obtained for sodium and potassium agree with the experimental results.
The thermodynamic coefficients (isothermal compressibility, density fluctuations, and thermal expansion) of sodium are investigated in the supercritical temperature region.
Numerical calculations of thermodynamic coefficients are performed close to the critical point, where carrying out theoretical and experimental research is challenging.
The change in compressibility with increasing density at various temperature values is traced.
The behavior of density fluctuations approaching the critical point is shown for different temperatures.
The variation in the magnitude of the thermal expansion with increasing temperature for different pressure values is illustrated.
}

\vspace{0.5cm}



\section{Introduction}
\label{sec:1}

Numerous works are devoted to explaining the critical properties of liquid systems. A detailed bibliography on the topic can be found, for example, in books \cite{hm113,gspmo110,amo191,bmo102} and review papers \cite{ss186,as100}.
Various theories represent the approaches for critical phenomena description in fluids.
In particular, the most common is the theory of integral equations, which consists of homogeneous Ornstein-Zernike equation \cite{hm113,k115} and the closure relations such as the Percus-Yevick \cite{py158} and hypernetted chain closures \cite{h103}. The fluctuation theory of solutions is actively used \cite{smomo113}.
The renormalization group (RG) theory \cite{wk174} also successfully describes the properties of systems near the critical point.
For example, Yukhnovskii used the conception of RG theory \cite{y118} to calculate the partition function of the Ising model and the grand partition function of a gas-liquid system in the collective variables phase space.
Computer simulation techniques also play a significant role in the study of critical phenomena and phase behavior of fluids \cite{b110}.

In the last decades, the interest in supercritical fluids, the study of their unique properties, and their application in different fields of science and technology has been growing steadily \cite{mg120,akm117,yl118,v118,pyy121}.
The reason for persistent interest in describing the nature of phase transitions and critical phenomena in liquid systems is that near-critical fluids are the most suitable objects for modeling a class of systems with many strongly interacting degrees of freedom \cite{gspmo110,amo191}.
On the other hand, the supercritical fluids are increasingly widely used in various technological processes due to their specific properties \cite{ztc106}.
In this regard, constructing the equation of state of supercritical fluids becomes a crucial applied problem.

One of the biggest challenges to successfully calculating the equation of state through statistical physics is accounting for the intricate interparticle interactions. In calculations, we use simplified models. They are limited in scope and determined for each specific case, either based on the internal properties of the model or by comparison with more accurate solutions or experimental results.

In this paper, we carried out the microscopic description and investigation of the fluid's critical behavior using a cell model within the framework of the grand canonical ensemble.
This issue is essential since the availability of chemical potential in the grand canonical ensemble leads to an adequate representation of existing atomic and molecular systems. Only this thermodynamic parameter is responsible for the exchange of constituents between different parts of the system and with the environment.
Moreover, it quantitatively describes the tendency of a thermodynamic system to establish compositional equilibrium.

The object of the present study is the Morse fluid in the supercritical region.

\section{Model}
\label{sec:2}

In this paper, a cell fluid model is used for studying the behavior of
a simple fluid in the vicinity of the liquid--gas critical point. We consider
a system of $N$ interacting particles in the volume $V$ conditionally
divided into $N_v$ cells ($V = vN_v$, $v = c^3$  is the cell volume,
and $c$ is the linear size of a cell) \cite{kd116,kpd118,p120,pk722}.
In contrast to a cell gas model (where a cell is assumed to contain only one particle or be empty)\cite{rebenko_13,rebenko_15} in our approach, a cell may hold more than one particle. Besides, the distance between the cells is introduced instead of the distance between the particles.

The grand partition function of the cell fluid model within the framework
of the grand canonical ensemble is as follows \cite{kpd118,p120}:
\be
\Xi = \sli_{N=0}^{\infty} \frac{(z)^N}{N!} \int \limits_{V} (dx)^N
\exp \left[-\frac{\beta}{2} \sli_{\vl_1,\vl_2\in\Lambda}
\tilde U_{l_{12}} \rho_{\vl_1} (\eta) \rho_{\vl_2} (\eta) \right].
\label{0d1fb2}
\ee
Here $z = e^{\beta \mu}$ is the activity, $\beta=1/(kT)$
is the inverse temperature, and $\mu$ is the chemical potential.
Integration with respect to coordinates of all the particles
$x_i = (x_{i}^{(1)},x_{i}^{(2)}, x_{i}^{(3)})$ is noted as
$\int \limits_{V} (dx)^N = \int \limits_{V} dx_1 \cdots \int \limits_{V} dx_N$,
and $\eta = \{ x_1 , \ldots , x_N \}$ is the set of coordinates.
The interaction potential $\tilde U_{l_{12}}$ is a function of the distance
$l_{12}= |\ve{l}_{1}-\ve{l}_{2}|$ between cells. Each vector $\ve{l}_i$ belongs
to the set
\be
\Lambda =\Big\{ \vl = (l_1, l_2, l_3)|l_i = c m_i;
m_i=1,2,\ldots,N_a; i=1,2,3; N_v=N_a^3 \Big\},
\label{jml2020-1fb2}
\ee
where $N_a$ is the number of cells along each axis. The occupation numbers of cells
\be
\rho_{\vl}(\eta) = \sli_{x \in \eta} I_{\Delta_{\vl}(x)}
\label{0d2fb2}
\ee
appearing in Eq. (\ref{0d1fb2}) are defined by
the characteristic functions (indicators)
\be
I_{\Delta_{\vl}(x)} = \left\{
\begin{array}{l}
1,\quad \texttt{if} \quad x\in\Delta_{\vl} \\
0,\quad \texttt{if} \quad x\notin\Delta_{\vl},
\end{array} \rd
\label{0d3fb2}
\ee
which identify the particles in each cubic cell
$\Delta_{\vl} = (-c/2,c/2]^3 \subset \mathbb{R}^3$ and their contribution to
the interaction of the model. Henceforth we choose the Morse potential as the interaction potential $\tilde U_{l_{12}}$:
\be
\tilde U_{l_{12}} = \Psi_{l_{12}} - U_{l_{12}}; \quad
\Psi_{l_{12}} = D e^{-2(l_{12}- 1)/\alpha_R}, \quad
U_{l_{12}} = 2 D e^{-(l_{12}- 1)/\alpha_R}.
\label{0d4fb2}
\ee
Here $\Psi_{l_{12}}$ and $U_{l_{12}}$ are the repulsive and attractive parts
of the potential, respectively, and $\alpha_R = \alpha / R_0$ ($\alpha$ is
the effective interaction radius). The parameter $R_0$ corresponds
to the minimum of the function $\tilde U_{l_{12}}$, and $D$ determines
the depth of a potential well. Note that the $R_0$-units are used for length measuring in terms of convenience. As a result,
$R_0$- and $R_0^3$-units are used for the linear size of each cell $c$ and
volume $v$, respectively.

\section{Calculational scheme and the equation of state}
\label{sec:3}

We recall the main idea of calculating the thermodynamic potential near the critical point within the approach of collective variables (CV) \cite{ymo287,YuKP_2001}. It consists of the separate inclusion of contributions from short-wave and long-wave modes of order parameter oscillations.
The short-wave modes are characterized by a RG symmetry and described by a non-Gaussian measure density.
They correspond to the critical regime region observed above and below the critical temperature $T_c$. In this case, the RG method is used.
We integrate the grand partition function of the cell fluid model over the layers of the CV phase space (see \cite{kpd118}).
The corresponding RG transformation can be related to the Wilson type.
Although, like the Wilson approach, the CV method exploits the RG ideas, it is based on the use of a non-Gaussian density of measure.
The main feature is the integration of short-wave oscillation modes, in general, without using perturbation theory.
As a result, we obtained the recurrence relations between the coefficients of the effective quartic measure densities, their solutions, and the equation for the critical temperature \cite{kpd118}.
Including short-wave oscillation modes leads to a renormalization of the dispersion of the distribution describing long-wave modes.
In the case of $T>T_c$, these long-wave modes correspond to the region of the limiting Gaussian regime.
We took into account the contribution from long-wave modes of oscillations to the thermodynamic potential of the cell fluid model in the way, which is qualitatively different from the method of calculating the short-wave part of the thermodynamic potential. The calculation of this contribution is based on the use of the Gaussian measure density as the basis one.
Here, we have developed a direct method of calculation with the results obtained by accounting for the short-wave modes as initial parameters.

\begin{figure}[htbp]
\centering \includegraphics[width=0.47\textwidth]{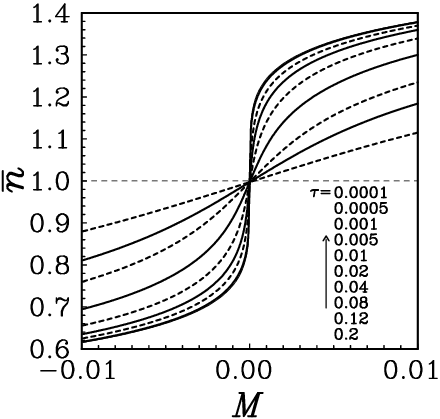}
\caption{Plot of the average density $\bar n$ as a function of
the chemical potential $M$ at various fixed values of the relative
temperature $\tau$. The arrow points on correspondence between $\tau$
and lines of $\bar n$ at the transition from bottom to top in the first
quadrant of the coordinate plane.}
\label{fig_1fa2}
\end{figure}

In the course of describing the behavior of the supercritical fluid, we obtained and investigated a nonlinear equation connecting the average density $\bar n$ and the quantity $M$, which is expressed by the initial chemical potential $\mu$ (hereafter, consider $M$ the chemical potential). This equation can be represented as \cite{kpd118}
\be
b_3^{(+)} M^{1/5}  = \bar n - n_g + M,
\label{3d41fa2}
\ee
where
\be
b_3^{(+)} = \lp \frac{b_1^{(+)}}{b_2^{(+)}} \rp^{1/5} \sigma_{00}^{(+)}, \quad
b_1^{(+)} = \lp \beta W(0) \rp^{1/2}, \quad
b_2^{(+)} = \frac{\alpha}{(1+\alpha^2)^{1/2}}.
\label{3d42fa2}
\ee
Here $W(0)$ is the Fourier transform of the effective interaction potential
at zero value of the wave vector, and $n_g$ is determined by the coefficients
of the initial expression for the grand partition function.
The coefficient $\sigma_{00}^{(+)}$ is a function of the quantity $\alpha$,
which is defined as the ratio of the renormalized chemical potential to
the renormalized temperature. The equation (\ref{3d41fa2}) allows for tracing the average density $\bar n$ as a function of $M$ ($M\ll 1$) at different fixed values of the relative temperature $\tau = (T-T_c)/T_c$ (see Fig.~\ref{fig_1fa2}).
Note that all the graphic
material represented in this paper is for the parameters of the Morse interaction
potential taken from \cite{p120}, which correspond to the data
for sodium \cite{singh}. Paper \cite{p120} also contains a set
of parameters for potassium. In particular, we have $R_0/\alpha = 2.9544$ for sodium
and $R_0/\alpha = 3.0564$ for potassium.

The expression for the logarithm of the grand partition function (or the thermodynamic potential $\Omega = -kT\ln \Xi$)
derived in \cite{kpd118} for the cell fluid model at $T>T_c$ makes it
possible to obtain the pressure $P$ as a function of the temperature
$T$ and the chemical potential $\mu$ using the well-known equation
\be
P V = kT \ln \Xi.
\label{3d2fa2}
\ee
Having the grand partition function, we can also find the average number
of particles
\be
\bar N = \frac{\partial \ln\Xi}{\partial \beta\mu}.
\label{3d3fa2}
\ee
The latter relation allows us to express the chemical potential in terms
of the average number of particles $\bar N$ or in terms of
the average density
\be
\bar n = \frac{\bar N}{N_v} = \lp \frac{\bar N}{V}\rp v,
\label{3d4fa2}
\ee
where $v$ is the volume of a cubic cell, which is the parameter of the model. Combining Eqs. (\ref{3d2fa2}) and (\ref{3d3fa2}), we
find the pressure $P$ as a function of the temperature $T$ and the average
density $\bar n$, in other words, the equation of state of our model.

\begin{figure}[htbp]
\centering \includegraphics[width=0.47\textwidth]{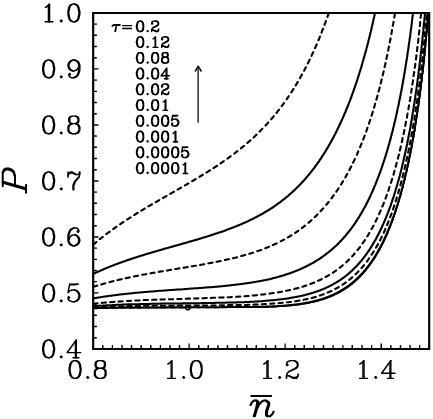}
\caption{Plot of the pressure $P$ as a function of the average density $\bar n$
at different fixed temperatures $\tau$. The critical point
($\bar n_c = 0.997$, $P_c = 0.474$) is marked by the symbol $\circ$.}
\label{fig_2fa2}
\end{figure}

The equation of state of the cell fluid model at $T>T_c$ obtained
using the simplest non-Gaussian quartic fluctuation distribution
(the $\rho^4$ model), takes the form \cite{kpd118}
\be
\frac{P v}{k T} =  P_a^{(+)}(T) + E_\mu +
\lp \frac{\bar n - n_g}{\sigma_{00}^{(+)}} \rp^6 \left[ e_0^{(+)}
\frac{\alpha}{(1+\alpha^2)^{1/2}} + \gamma_s^{(+)} - e_2^{(+)} \right].
\label{3d47fa2}
\ee
The quantity $P_a^{(+)}(T)$ appearing in Eq. (\ref{3d47fa2}) contains
an analytical dependence on temperature. The coefficient $\gamma_s^{(+)}$
characterizes the nonanalytic contribution to the thermodynamic potential.
The quantities $e_0^{(+)}$, $e_2^{(+)}$, and $\sigma_{00}^{(+)}$
depend on the roots of a specific cubic equation. The expressions for all
these quantities, as well as for $E_{\mu}$, are given in \cite{kpd118}.
Using Eq. (\ref{3d47fa2}), in Fig.~\ref{fig_2fa2}, we demonstrate the pressure behavior as the density increases.

\section{Results}
\label{sec:4}

The proposed analytical approach, developed to describe the critical behavior of the cell fluid model by taking into account non-Gaussian fluctuations of the order parameter (the $\rho^4$ model approximation), is applied to the Morse potential parameters characteristic of simple alkali metals (sodium and potassium).
The critical point temperature $kT_c$ can be calculated using the equation obtained in our paper \cite{kpd118}.
Eqs. (\ref{3d41fa2}) and (\ref{3d47fa2}) give expressions for the critical fluid density $\bar n_c$ and pressure $P_c$, respectively.
Table~\ref{tab_1fb2} shows numerical estimates of the critical point parameters, which we obtained for sodium and potassium. The results of the so-called zero mode approximation (the mean-field approximation) (see \cite{kdp117}) are in the first row, and those based on the proposed theory (see \cite{kpd118,p120} and this paper) are in the second row. For comparison, Table~\ref{tab_1fb2} also contains the results of other authors.
\begin{table}[htbp]
\caption{Dimensionless critical temperature $kT_c$, average density $\bar n_c$,
and pressure $P_c$, obtained for sodium (Na) and potassium (K)
on the basis of the cell fluid model (CFM) in the zero mode approximation
(ZMA) \cite{kdp117} (see the first row) and in the $\rho^4$ model approximation
(R4MA) \cite{kpd118,p120} (see the second row). The third and fourth rows of
the table refer to the Monte Carlo simulation results \cite{singh} and
experimental data \cite{hensel}, respectively.}
\label{tab_1fb2}
\begin{center}
\begin{tabular}{cccccccc}
\hline
\multicolumn{1}{c}{Research methods} & \multicolumn{3}{c}{Na} &
\multicolumn{4}{c}{K} \\
\cline {2-4}
\cline {6-8}
& \multicolumn{1}{c}{$kT_c$} & \multicolumn{1}{c}{$\bar n_c$} & \multicolumn{1}{c}{$P_c$}
& & \multicolumn{1}{c}{$kT_c$} & \multicolumn{1}{c}{$\bar n_c$} & \multicolumn{1}{c}{$P_c$} \\
\hline
\multicolumn{1}{l}{Theory\,(CFM,\,ZMA)} & 5.760 & 0.997 & & & 5.037 & 0.935 & \\
\multicolumn{1}{l}{Theory\,(CFM,\,R4MA)} & 4.028 & 0.997 & 0.474 & & 3.304 & 0.935 & 0.408 \\
\multicolumn{1}{l}{Simulations}
& 5.874 & 1.430 & 2.159 & & 5.050 & 1.125 & 1.651 \\
\multicolumn{1}{l}{Experiment} & 3.713 & 1.215 & 0.415 & & 3.690 & 0.772 & 0.498 \\
\hline
\end{tabular}
\end{center}
\end{table}

The zero mode approximation does not account for the fluctuations of the order parameter. This approximation is inefficient near the critical point, where fluctuation effects play a significant role. The $\rho^4$ model approximation takes into account the non-Gaussian order parameter fluctuations that lead to the emergence of a renormalization group symmetry. Table~\ref{tab_1fb2} allows evaluation of the discrepancies between theoretical, experimental, and Monte Carlo simulation results. As can be seen from the Table~\ref{tab_1fb2}, our estimates of the critical point parameters for Na and K in the $\rho^4$ model approximation agree better with the experimental data \cite{hensel} than the numerical results \cite{singh} obtained by Monte Carlo simulations.
The critical temperatures for Na and K are overestimated in Monte Carlo calculations. The critical pressures for the alkali metals show significant deviations from theoretical and experimental values. For Na, the pressure is vastly overestimated because the critical temperature is overvalued. It is observed that at the experimental critical point of Na metal, 2485 K, the corresponding pressure predicted by simulation is a good approximation to the experimental critical pressure.
In \cite{singh}, the authors note that the critical properties of Na and K are overestimated by their simulations, which means that the used parameters need to be refined to give a better agreement with experimental data. Scaling of the parameters to correctly predict the literature values, which are also observed to have a wide scatter, is reserved for further study.

Using the equation of state (\ref{3d47fa2}), we can trace the variation
of the fluid density with increasing temperature for
different pressure values, calculate and investigate thermodynamic coefficients (isothermal compressibility,
density fluctuations, and thermal expansion) in
the supercritical temperature region ($T>T_c$).

\begin{figure}[htbp]
\includegraphics[width=0.47\textwidth]{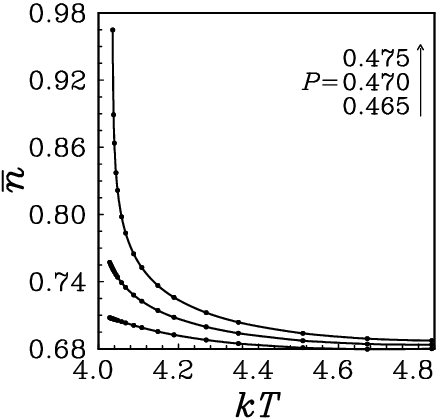}
\hfill
\includegraphics[width=0.47\textwidth]{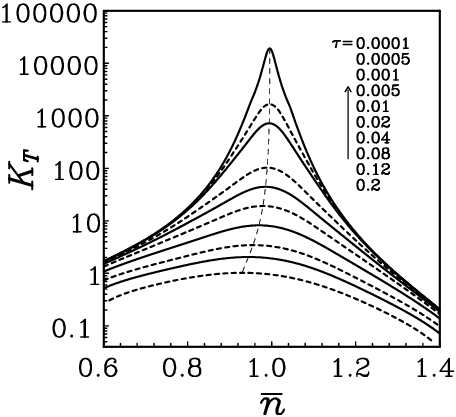}
\\
\parbox[t]{0.47\textwidth}{
\vspace{-0.3cm}
\caption{Plot of the average density $\bar n$ as a function of temperature
for various fixed values of the pressure $P$.}
\label{fig_3fa2}
}
\hfill
\parbox[t]{0.47\textwidth}{
\vspace{-0.3cm}
\caption{Evolution of the isothermal compressibility as the
density increases at fixed values of the relative temperature.}
\label{fig_4fa2}
}
\end{figure}

\begin{figure}[htbp]
\includegraphics[width=0.47\textwidth]{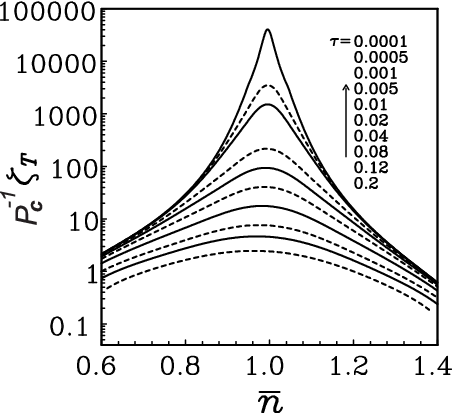}
\hfill
\includegraphics[width=0.47\textwidth]{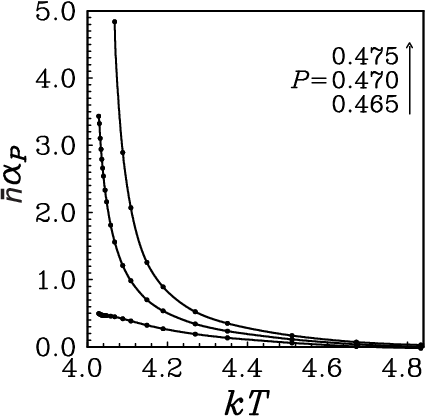}
\\
\parbox[t]{0.47\textwidth}{
\vspace{-0.3cm}
\caption{Density fluctuations for tempe\-ratures close to $T_c$.}
\label{fig_5fa2}
}
\hfill
\parbox[t]{0.47\textwidth}{
\vspace{-0.3cm}
\caption{Temperature dependence of the thermal expansion $\alpha_P$
for various fixed values of the pressure $P$.}
\label{fig_6fa2}
}
\end{figure}

The subsequent figures illustrate the results of our numerical calculations
obtained for sodium in the vicinity of the critical point.
The temperature dependence of the average density at various fixed
pressure values is shown in Fig.~\ref{fig_3fa2}.
Fig.~\ref{fig_4fa2} demonstrates the density dependence of the isothermal
compressibility
\be
K_T = \frac{1}{\eta} \lp \frac{\partial \eta}{\partial \tilde p}\rp_T =
\frac{P_c}{\bar n} \lp \frac{\partial P}{\partial \bar n}\rp_T^{-1}
\label{zv2022-1fa2}
\ee
for fixed temperature values (we used the same values of $\tau$
to plot the isotherms in Fig.~\ref{fig_2fa2}) \cite{kpd118}.
Here $\eta = \bar n/\bar n_c$, $\tilde p = P/P_c$.
Proceeding from the extreme values of the isothermal compressibility $K_T$
(the dashed line in Fig.~\ref{fig_4fa2}), we have constructed the Widom
line of a supercritical cell fluid in \cite{kpd118}.
The density fluctuations
\be
\zeta_T = t \lp \frac{\partial \eta}{\partial \tilde p}\rp_T =
\frac{P_c (1 + \tau)}{\bar n_c} \lp \frac{\partial P}{\partial \bar n}\rp_T^{-1}
\label{sem2020-1fa2}
\ee
and the thermal expansion coefficient
\be
\alpha_P = - \frac{1}{\eta} \lp \frac{\partial \eta}{\partial t}\rp_P =
- \frac{1}{\bar n} \lp \frac{\partial \bar n}{\partial \tau}\rp_P,
\label{sem2020-2fa2}
\ee
calculated in addition to the isothermal compressibility, are shown
in Figs.~\ref{fig_5fa2} and \ref{fig_6fa2}, respectively. Here $t = T/T_c$.
The graphs in Figs.~\ref{fig_4fa2} and \ref{fig_5fa2} are similar to each
other, since the behavior of $K_T$ and $\zeta_T$
is determined by the same derivative $(\partial P/\partial \bar n)_T$
[see Eqs. (\ref{zv2022-1fa2}) and (\ref{sem2020-1fa2})].

Note that this calculation is valid in a small neighborhood of $T_c$, which is problematic for theoretical and experimental studies.
The solutions of recurrence relations (see \cite{kpd118}) allow us to calculate the size of the critical region. In these solutions, the terms proportional to $E_3^n$
describe the entry to the critical regime, and the terms
proportional to $E_2^n$ describe the exit from the critical regime. Here $E_2$ and $E_3$ are the eigenvalues of the RG linear transformation matrix, and $n$ is the layer number in the CV phase
space. We can determine the temperature range $\tau < \tau^*$ in which
the critical regime emerges using the solutions of recurrence relations and the condition
for the critical regime existence (the exit from the critical regime
for $n \rightarrow 1$ does not prevail over the entry to this
regime). The temperature $\tau^*$ equals the magnitude
of the smallest root of two equations obtained from
the recurrence relations solutions. The quantity $\tau^*$ determined
in this way is of the order of a few hundredths ($\tau^* = 0.04$
in the case of liquid sodium and $\tau^* = 0.02$ for potassium).
The region of interest for most applications of supercritical fluids
covers this temperature value (usually $1 < T/T_c < 1.1$
(or $0 < \tau < 0.1$) and $1 < P/P_c < 2$ \cite{ekd196}).

\section{Conclusions}
\label{sec:5}

In this paper, we studied the behavior of the fluid system within the cell model in the immediate vicinity of the critical point. The region close to the critical point is attractive because of the fundamental and applied aspects. At the same time, it is complicated to analyze due to the essential role of fluctuation effects.

We developed the method for Morse fluids and applied it to describe a phase transition in simple liquid alkali metals. The parameters of the Morse interaction potential used for calculations are inherent to alkali metals (sodium and potassium).
We have calculated the critical point parameters for sodium and potassium, which are in agreement with the experimental data. Through our developed approach that considers non-Gaussian fluctuations of the order parameter, we have obtained the equation of state of the cell fluid model, which gives the pressure as a function of temperature and density.
We use this equation to study the thermodynamic coefficients of
Morse fluids. Thermodynamic coefficients (isothermal compressibility,
density fluctuations, and thermal expansion) are calculated in the case
of sodium. We demonstrate the density fluctuations and thermal expansion in the supercritical temperature region, besides the isothermal compressibility obtained previously.

Fig.~\ref{fig_2fa2} shows that the isotherms acquire a flat shape close to the critical point, i.e., the slope ($\partial P/\partial \bar n$) goes to zero at
$T\longrightarrow  T_c^+$. This corresponds to the fact that the isothermal compressibility (\ref{zv2022-1fa2}) and the density fluctuations (\ref{sem2020-1fa2}) become enormous when approaching the critical point (see Figs.~\ref{fig_4fa2} and \ref{fig_5fa2}). Very large values of the isothermal compressibility mean a substantial density sensitivity to tiny pressure fluctuations.

The behavior of the temperature-dependent average density and the thermal expansion at the fixed pressures is shown in Figs.~\ref{fig_3fa2} and \ref{fig_6fa2}. For various fixed pressure, we see a decrease in the thermal expansion coefficient (\ref{sem2020-2fa2}) with increasing supercritical temperature, which is typical of gases. The thermal expansion coefficient of gases with increasing temperature approaches the value of the thermal expansion coefficient of an ideal gas, which is equal to the reciprocal absolute temperature.

Our approach to simple fluid systems will help to study the critical behavior of multicomponent fluids. Our research provides a particular methodological contribution to the theoretical description of critical phenomena.

\end{document}